\documentclass[10pt]{iopart}
\usepackage{epsfig,cite}
\newcommand{\ewxy}[2]{\setlength{\epsfxsize}{#2}\epsfbox[30 30 640 640]{#1}}
\newcommand{\beq}{\begin{equation}}
\newcommand{\eeq}{\end{equation}}
\newcommand{\ds}{\displaystyle}
\newcommand{\beqar}{\begin{eqnarray}}
\newcommand{\eeqar}{\end{eqnarray}}

\begin{document}

\title[Flow of $K$'s and $\Lambda$'s in HICs at RHIC]{Anisotropic 
flow of strange particles in heavy ion collisions at RHIC energies}
\author{
L~V~Bravina\dag\ddag, L~P~Csernai\S, Amand~Faessler\dag, 
C~Fuchs\dag, E~E~Zabrodin\dag\ddag
}
\address{\dag\
Institut f\"ur Theoretische Physik, Universit\"at
    T\"ubingen, T\"ubingen, Germany}
\address{\ddag\
Institute for Nuclear Physics, Moscow State University, Moscow, 
    Russia}
\address{\S\
Department of Physics, University of Bergen, Norway}

\begin{abstract}
Anisotropic flow of $K$'s, $\overline{K}$'s, and $\Lambda$'s is
studied in heavy ion collisions at SPS and RHIC energies within the 
microscopic quark-gluon string model. At SPS energy the directed 
flow of kaons differs considerably at midrapidity from that of 
antikaons, while at RHIC energy kaon and antikaon flows coincide.
The change is attributed to formation of dense meson-dominated matter 
at RHIC, where the differences in interaction cross-sections of 
hadrons become unimportant. The directed flows of strange particles,
$v_1^{K,\overline{K}, \Lambda}(y)$, have universal negative slope at 
$|y| \leq 2$ at RHIC. 
The elliptic flow of strange hadrons is developed at midrapidity at 
times $3 \leq t \leq 10$ fm/$c$. It increases almost linearly with 
rising $p_t$ and stops to rise at $p_t \geq 1.5$ GeV/$c$ reaching the 
same saturation value $v_2 ^{K,\Lambda}(p_t) \approx 10\%$ in accord 
with experimental results.
\end{abstract}

\section{Introduction}
\label{intro}
The transverse collective flow of particles is
at present one of the most intensively studied characteristics of
heavy-ion collisions \cite{flow}, because the flow
is directly linked to the equation of state (EOS) of the system.
If even a small amount of the quark-gluon plasma (QGP) is formed in 
the course of the collision, it would lead to a reduction of pressure
and a softening of the EOS that can be detected experimentally.
To study the properties of transverse particle flow the method of
Fourier series expansion \cite{VoZh96} has been proved to
be very useful:
\beq
\ds
E \frac{d^3 N}{d^3 p} = \frac{d^2 N}{2 \pi p_t dp_t dy} \left[
1 + 2 \sum_{n=1}^{\infty} v_n \cos(n\phi) \right] .
\label{eq1}
\eeq
Here $p_t$, $y$, and $\phi$ are the transverse momentum, rapidity,
and the azimuthal angle of a particle, respectively.
The unity in square brackets represents the isotropic radial
flow, while the other terms are refer to anisotropic, directed
$v_1=\langle \cos{\phi} \rangle$ and elliptic $v_2 = \langle
\cos{(2 \phi)} \rangle$, flow.
The idea that the elliptic flow can carry important information
about the early stage of heavy-ion collisions has been discussed
first in Ref.~\cite{Olli92}. This suggestion is supported by
macroscopic hydrodynamic and microscopic transport simulations,
which show that elliptic flow saturates quite early
\cite{Sorprl97,HeLe99,ZGK99,KHHH01}, while directed flow
develops almost until the stage of final interactions
\cite{LPX99}. However, the directed flow of hadrons with high
transverse momentum can be used as a probe of hot and dense phase of
the collision \cite{trflprc01} due to the early freeze-out times of
these particles.

In the present paper we are studying the anisotropic flow of kaons,
antikaons, and lambdas produced in lead-lead and gold-gold collisions
at $E_{lab} = 160$ AGeV (SPS) and $\sqrt{s} = 130$ AGeV (RHIC), 
respectively.
For the study the microscopic cascade quark-gluon string model (QGSM)
\cite{qgsm}, based on Gribov-Regge theory accomplished by a string 
phenomenology of particle production in inelastic hadron-hadron 
collisions, is employed. The model successfully describes the 
elliptic flow of charged particles at RHIC \cite{plb01} measured 
recently \cite{ell_fl_rhic}, as well as other characteristics.

\section{Anisotropic flow at SPS and RHIC}
\label{flow}
\begin{figure}[htb]
\vspace{-4.0cm}
\hspace{1.cm}
\ewxy{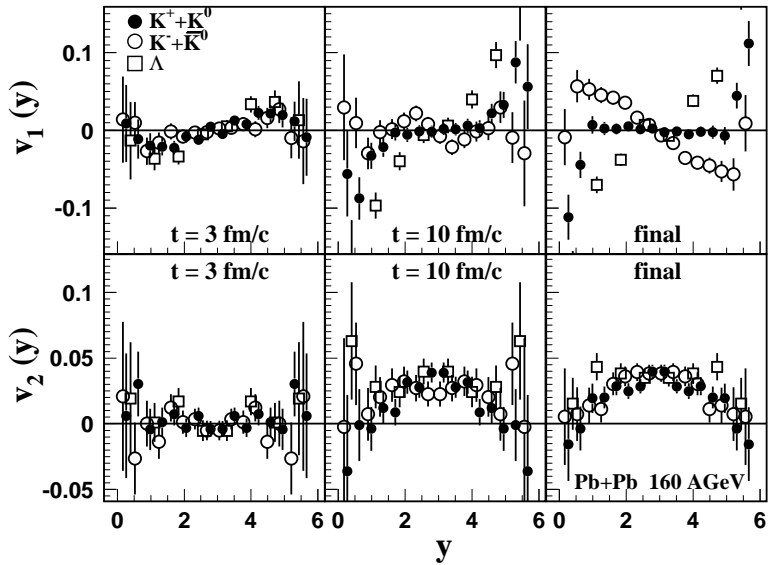}{118mm}  
\vspace{-0.6cm}
\caption{
Time evolution of directed $v_1(y)$ and elliptic $v_2(y)$ flow of 
kaons, antikaons, and lambdas in minimum bias Pb+Pb collisions at SPS.
}
\label{fig1}
\end{figure}

Time evolution of directed and elliptic flow of strange particles in 
minimum bias Pb+Pb collisions at SPS is shown in Fig.~\ref{fig1}. Here 
the anisotropic flow is calculated at early, $t=3$ fm/$c$ and $t = 10$
fm/$c$, and at the final stage of the reaction. To avoid ambiguities,
all resonances in the scenario with early freeze-out were allowed to 
decay according to their branching ratios. At early stages of the
collision directed flow of all strange particles is oriented in the
direction of normal flow similar to that of nucleons. At this stage
the matter is dense, mean free paths of particles are short,
and similarities in hadron production and rescattering dominate over 
inequalities caused by different interaction cross-sections.
Later on the system becomes more dilute. Due to larger interaction
cross-sections of antikaons with other hadrons, the directed flow of 
these particles changes the orientation from a weak normal to strong 
antiflow.

Elliptic flow of $K$'s, $\overline{K}$'s and $\Lambda$'s is close to
zero at $t=3$ fm/$c$, i.e., the particle distribution is isotropic in 
the transverse plane. The resulting elliptic flow of strange particles
at SPS is positive. This feature can be explained
by secondary rescattering in the spatially anisotropic matter that
lead to increase of the elliptic flow along the impact parameter
axis. Note that at midrapidity the elliptic flow is formed at 
$t \leq 10$ fm/$c$, that quantitatively agrees with the estimates 
made in \cite{Olli92}.

\begin{figure}[htb]
\vspace{-4.0cm}
\hspace{1.cm}
\ewxy{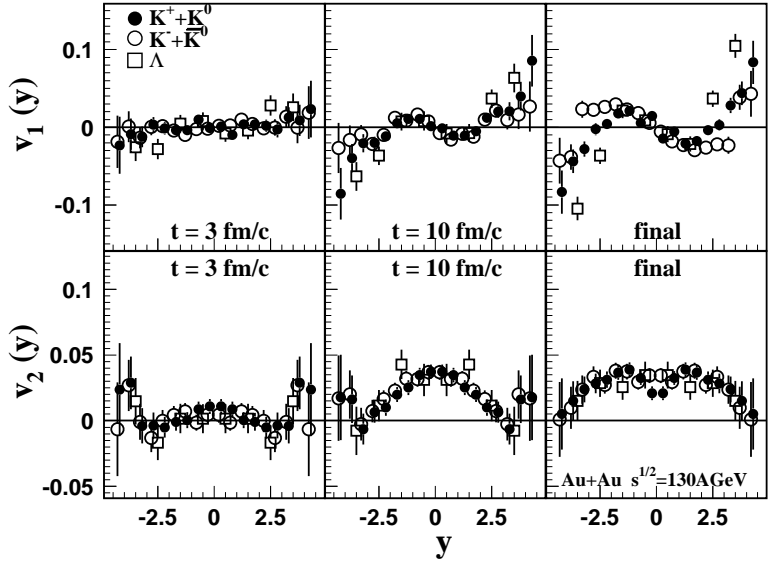}{118mm}  
\vspace{-0.6cm}
\caption{
The same as Fig.~\ref{fig1} but for Au+Au collisions at RHIC.
}
\label{fig2}
\end{figure}

Time evolution of the anisotropic flow of strange particles at RHIC
is depicted in Fig.~\ref{fig2}. At midrapidity directed flow of 
$K$'s, $\overline{K}$'s, and $\Lambda$'s is nearly zero at $t = 3$
fm/$c$, but it changes to strong antiflow at $t \geq 10$ fm/$c$
with the similar slopes for all strange hadrons. The effect is 
traced to nuclear shadowing which can mimic formation of the QGP 
(see \cite{shad}) at ultra-relativistic energies. Final elliptic flow
of all particles clearly shows the in-plane orientation. Directed and 
elliptic flows of $K$'s and $\overline{K}$'s coincide within the 
statistical error bars. This is attributed to the formation of dense 
baryon-dilute matter, where the difference in interaction cross 
sections of particles play a minor role.

\begin{figure}[htb]
\vspace{-2.5cm}
\hspace{1.cm}
\ewxy{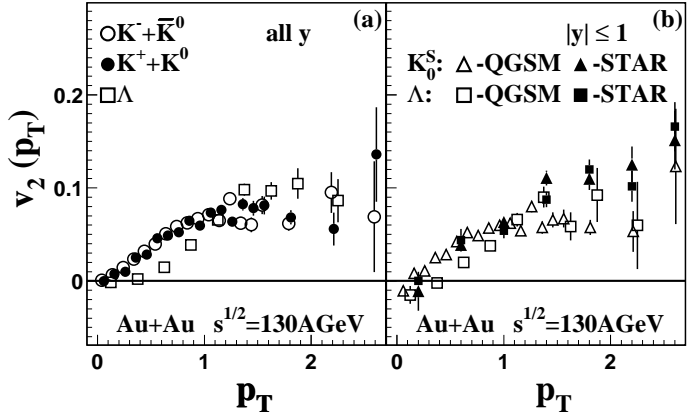}{98mm}  
\vspace{-0.6cm}
\caption{
Transverse momentum dependence of the elliptic flow of strange
particles in the whole rapidity (a) and in central rapidity interval 
(b) in minimum bias Au+Au collisions at RHIC. Data taken from
\protect\cite{starKL}.
}
\label{fig3}
\end{figure}

Figure~\ref{fig3} presents the $p_T$-dependence of the elliptic flow
of strange hadrons at RHIC. The flow is close to zero for hadrons
with $p_T \leq 0.25\,$GeV/$c$, then rises linearly up to $v_2^K (p_T) 
\approx 10\%$ irrespective of the hadron mass within the interval 
$0.25 \leq p_T \leq 1.5\,$GeV/$c$, and saturates at $p_T \geq 
1.5\,$GeV/$c$ in accord with the experimental data \cite{starKL}. 
This behaviour can be explained by the interplay between the flow of 
high-$p_T$ particles, emitted at the onset of the collision, and the 
hydro-type flow of particles, which gained their high $p_T$ in 
secondary interactions.

\section{Conclusions}
\label{concl}
The results may be summarised as follows.
(1) The directed flow of kaons $v_1^K(y)$ produced in
minimum bias Pb+Pb collisions at SPS is found
to be close to zero in the midrapidity range, while the 
$v_1^{\overline{K}}(y)$ has a linear antiflow slope because of
different interaction cross sections and large absorption cross 
section of antikaons with baryons.
(2) In heavy ion collisions at RHIC a dense meson-dominated matter 
is produced, and the directed flow of $K$'s becomes similar to that
of $\overline{K}$'s. 
(3) The directed flows of $K$'s, $\overline{K}$'s, and $\Lambda$'s 
have universal negative slope at $|y| \leq 2$ at RHIC.
(4) The elliptic flow of strange particles is built up at 
midrapidity at $3 \leq t \leq 10$ fm/$c$ both at SPS and at RHIC.
At RHIC it increases linearly with rising $p_T$ and reaches 
saturation value $v_2^{K,\overline{K},\Lambda}(p_T) \approx 10\%$
at $p_T \geq 1.5\,$GeV/$c$.

{\bf Acknowledgements.} Discussions with J.-Y. Ollitrault,
S. Panitkin, and N. Xu are gratefully acknowledged.
L.B. acknowledges financial support from the A. v. Humboldt
Foundation.
This work was supported in part by the BMBF under contract No.
06T\"U986 and by the BCPL in the framework of the European Community 
- Access to Research Infrastructure action of the Improving Human 
Potential Programme.

\end{document}